\documentclass{IEEEtran}
\usepackage{cite}
\usepackage{amsmath,amssymb,amsfonts}
\usepackage{graphicx}
\usepackage{tabularx}
\usepackage{booktabs}
\usepackage[skip=0pt, size=footnotesize]{caption}

\newcolumntype{Y}{>{\centering\arraybackslash}X}

\begin{document}
\newpage
\title{Multi-Subject Image Synthesis as a Generative Prior for Single-Subject PET Image Reconstruction}
\author{George Webber, Yuya Mizuno, Oliver D. Howes, Alexander Hammers, Andrew P. King and Andrew J. Reader
\thanks{\scriptsize Presented at IEEE NSS MIC RTSD 2024, see doi: 10.1109/NSS/MIC/RTSD57108.2024.10657446. Copyright (c) 2024 IEEE.}
\thanks{\scriptsize George Webber would like to acknowledge funding from the EPSRC Centre for Doctoral Training in Smart Medical Imaging [EP/S022104/1] and via a GSK Studentship.}
\thanks{\scriptsize George Webber (e-mail: george.webber@kcl.ac.uk), Andrew P. King and Andrew J. Reader are with the School of Biomedical Engineering and Imaging Sciences, King’s College London, UK. Yuya Mizuno and Oliver D. Howes are with the Institute of Psychiatry, Psychology and Neuroscience, King’s College London, UK. Alexander Hammers is with King’s College London and Guy's \& St Thomas' PET Centre.}
}
\maketitle

\begin{abstract}
Large high-quality medical image datasets are difficult to acquire but necessary for many deep learning applications.
For positron emission tomography (PET), reconstructed image quality is limited by inherent Poisson noise.
We propose a novel method for synthesising diverse and realistic pseudo-PET images with improved signal-to-noise ratio.
We also show how our pseudo-PET images may be exploited as a generative prior for single-subject PET image reconstruction.
Firstly, we perform deep-learned deformable registration of multi-subject magnetic resonance (MR) images paired to multi-subject PET images.
We then use the anatomically-learned deformation fields to transform multiple PET images to the same reference space, before averaging random subsets of the transformed multi-subject data to form a large number of varying pseudo-PET images.
We observe that using MR information for registration imbues the resulting pseudo-PET images with improved anatomical detail compared to the originals.
We consider applications to PET image reconstruction, by generating pseudo-PET images in the same space as the intended single-subject reconstruction and using them as training data for a diffusion model-based reconstruction method.
We show visual improvement and reduced background noise in our 2D reconstructions as compared to OSEM, MAP-EM and an existing state-of-the-art diffusion model-based approach.
Our method shows the potential for utilising highly subject-specific prior information within a generative reconstruction framework.
Future work may compare the benefits of our approach to explicitly MR-guided reconstruction methodologies.
\end{abstract}

\vspace{-0.3cm}

\vspace{-0.5cm}
\section{Introduction}
\label{sec:introduction}
The quality of reconstructed positron emission tomography (PET) images is limited by the number of counts recorded.
Recent state-of-the-art results on PET reconstruction tasks have been achieved by utilising score-based generative models (SGMs), also known as diffusion models, as generalised inverse problem solvers \cite{singh_score-based_2024}.

The usefulness of a deep-learned image manifold as a prior for a reconstruction problem is constrained by the quantity, quality and relevance of the training data used to learn it.

We consider registering and summing inter-subject PET volumes, to obtain diverse reference data with higher signal-to-noise ratio. We then investigate training an SGM with this data to learn a high quality, highly relevant prior for PET image reconstruction.



\begin{figure*}
    \vspace{-0.9cm}
    \centering
    \begin{minipage}[t]{0.54\textwidth}
        \centering
        \includegraphics[width=\textwidth]{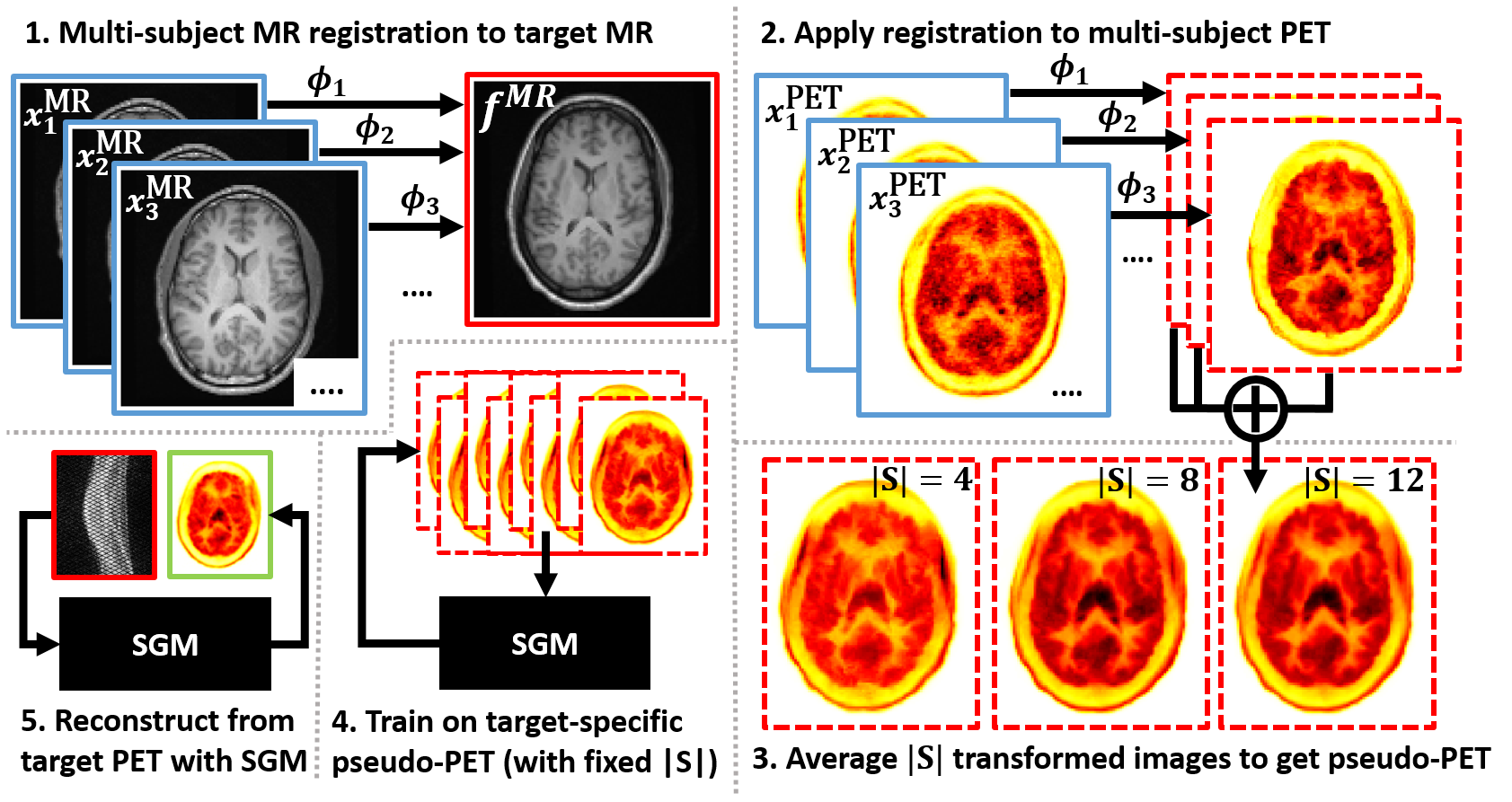}
        \caption{\footnotesize Outline of our reconstruction methodology: generating pseudo-PET images in our single-subject target space; incorporating target-specific pseudo-PET images as a prior into SGM-based reconstruction; and, reconstructing from target PET data.}
        \label{fig:methodology}
    \end{minipage}\hfill
    \begin{minipage}[t]{0.44\textwidth}
        \centering
        \includegraphics[width=\textwidth]{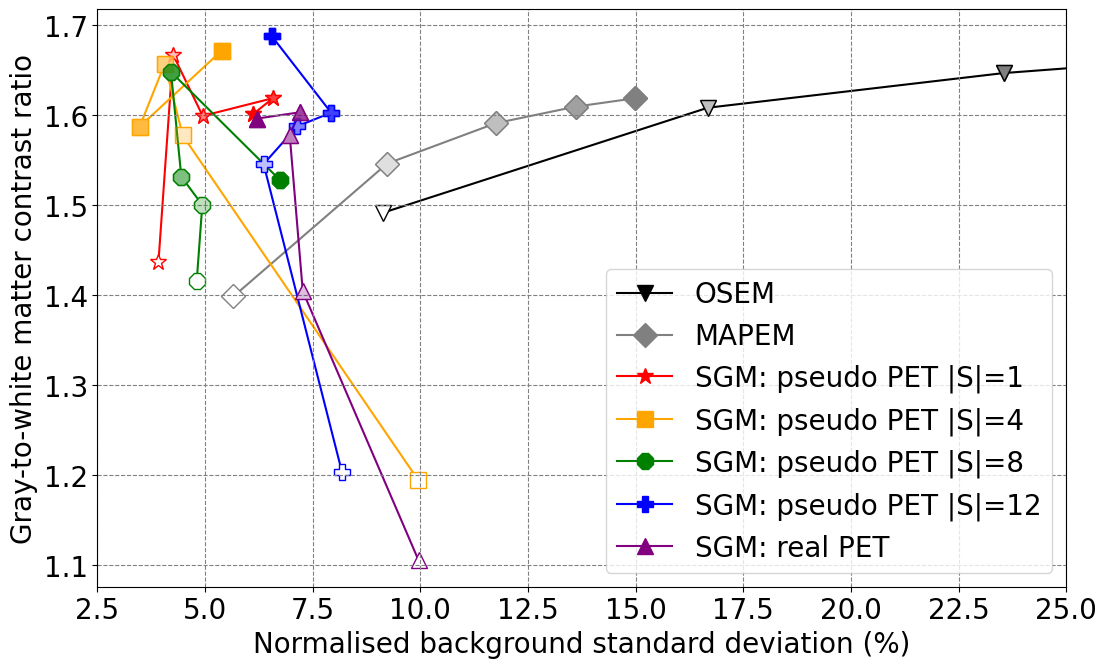} 
        \caption{\footnotesize Gray-to-white matter contrast ratio plotted against normalised standard deviation in a background region of white matter, for iteration number 20 (unfilled) to 100 (filled).}
        \label{fig:mean_roi}
    \end{minipage}
\end{figure*}

\section{Theory}


SGMs are a deep learning framework that may be employed to learn image distributions, given training samples from the distribution. To sample new images from the learned distribution, an SGM samples a standard Gaussian and repeatedly applies its learned de-noising operator. We can leverage an SGM as a learned prior for PET image reconstruction by interleaving the de-noising generative steps with steps that encourage consistency with measured PET sinograms.

Image registration is the problem of finding an optimal transformation from a ``moving'' image to a ``fixed'' image.
For affine registration, given fixed image $\mathbf{f}$ and moving image $\mathbf{x}$, we seek the affine map $A$ that minimises the loss $\mathcal{L}(\mathbf{f}, A\mathbf{x})$.
More generally, we seek a well-regularised deformation field $\phi$ that minimises $\mathcal{L}(\mathbf{f}, \mathbf{x} \circ \phi)$.
VoxelMorph \cite{balakrishnan_voxelmorph_2019} is a deep learning framework that parameterises the deformation field by its inputs $ \mathbf{f} $ and $ \mathbf{x} $, such that we seek to learn $\phi(\cdot, \cdot)$ given a training dataset. The deformation function $\phi(\cdot, \cdot)$ is based on a U-Net architecture, and outputs the deformation field between inputs $ \mathbf{f} $ and $ \mathbf{x} $. Its training loss sums a voxel-wise registration error $ || \mathbf{f}- \mathbf{x} \circ \phi(\mathbf{f},\mathbf{x}) ||_2^2$ and a smoothness penalty on $\phi(\mathbf{f},\mathbf{x})$.

Suppose we have a deformation function $ \phi(\cdot, \cdot) $ learned from magnetic resonance (MR) images as well as $M$ input training paired PET-MR images $\{(\mathbf{x}^{\text{PET}}_i, \mathbf{x}^{\text{MR}}_i)\}_{i=1}^{M}$. To construct a pseudo-PET image, we firstly define a reference space by choosing a reference MR image $\mathbf{f}^{\text{MR}}$. We then transform each PET image $\mathbf{x}^\text{PET}_i$ to the reference space by application of the MR deformation field $ \phi(\mathbf{f}^{\text{MR}}, \mathbf{x}^\text{MR}_i) $. We finally average a randomly-chosen subset $S$ of the transformed images.


The method offers the natural extensions of varying $ \mathbf{f}^{\text{MR}} $ and taking randomly weighted averages of PET images to generate an arbitrarily large dataset.
In this work, we fix $ \mathbf{f^{\text{MR}}} $ as the MR image of the subject whose PET data is to be reconstructed (our ``target'' PET data).
We then choose a fixed $|S|$, and use random pseudo-PET images (sampled without replacement) as highly subject-specific training data for an SGM reconstruction methodology based on PET-DDS (PET-Decomposed Diffusion Sampling), recently proposed by Singh et al. \cite{singh_score-based_2024}. See Figure \ref{fig:methodology} for an overview of our methodology.

\begin{figure*}[t]
    \vspace{-0.4cm}
    \centering
    \begin{center}
    	\includegraphics[width=\textwidth]{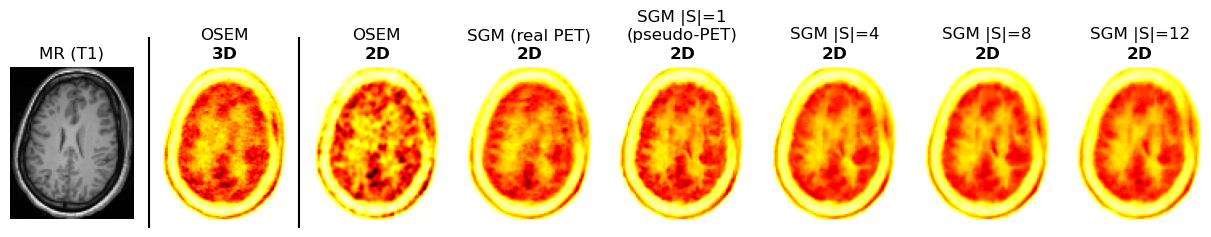}
    \end{center}
    \vspace{-0.2cm}
	\caption{\footnotesize One transverse brain slice reconstructed with different variants of our methodology, with baseline images for comparison. OSEM 3D is reconstructed from fully 3D PET sinograms, while other PET images are reconstructed from a single 2D direct sinogram.}
    \label{fig:reconstruction}
    \vspace{-0.5cm}
\end{figure*}

\vspace{-0.2cm}
\section{Experiments}
Thirteen static $[^{18}$F]DPA-714 brain datasets (from the Inflammatory Reaction in Schizophrenia team at King's College London) are used in this work \cite{muratib_dissection_2021}. Data were acquired from 1-hour scans with the Siemens Biograph mMR, with approximately 200 MBq administered, with total counts in the range $6.0 - 9.3 \times 10^8$. Standard reconstructed PET images (voxel size 2 $\times$ 2 $\times$ 2 $\text{mm}^3$ ; 3D image size of $344 \times 344 \times 127$) were formed using the default OSEM reconstruction with 2 iterations and 21 subsets. MR images were obtained with the MP-RAGE sequence (voxel size 1 $\times$ 1 $\times$ 1 $\text{mm}^3$ ; 3D image size of $224 \times 256 \times 176$).

MR images were firstly resampled to the same voxel size as the 3D PET images. Affine registration to a single reference MR scan was then performed with NiftyReg, before images were cropped to $128 \times 128 \times 120$. Next, VoxelMorph's deformation function was trained using the pre-processed MR volumes. Registration only on the PET images was investigated but found to be of poor quality, due to the relative lack of anatomical detail in the $[^{18}$F]DPA-714 images.

Using our learned deformation function and the fixed MR image for our reconstruction target, we held $|S|$ constant and generated 3D pseudo-PET images by summing random subsets of either 1, 4, 8 or 12 transformed PET images. The first 12 PET datasets were used in this task, with 1 reserved for testing.

For each value of $|S|$, an SGM was trained on shuffled batches of 2D transverse image slices taken from 2400 (non-unique) pseudo-PET 3D images. For comparison, an SGM was also trained on the original PET images.

Reconstruction was performed in 2D, using real prompt data from 127 2D direct transverse PET slices as well as scanner-calculated contamination and correction sinograms. PET-DDS was chosen as the SGM-based reconstruction method \cite{singh_score-based_2024}. Baseline methods of OSEM and MAP-EM (with patch-based regularisation \cite{wang_penalized_2012}) were implemented in 2D for comparison.

\vspace{-0.1cm}
\section{Preliminary Results and Discussion}

Considering Figure \ref{fig:methodology} parts 2 \& 3, we can see that the registered pseudo-PET images have gained anatomical information from the MR deformation field. Additionally, they are less noisy than real PET images (more so as $|S|$ increases).


For our reconstruction task, Figure \ref{fig:reconstruction} shows that our pseudo-PET images have a strong impact on our final reconstructions. Our 2D reconstructions surpass even the OSEM 3D reconstruction for visual fidelity (which has $\sim 8 \times$ more counts).

The number of unique training examples with $|S|$ fixed is $^{12}C_{|S|}$. Hence our $|S|=12$ SGM was trained on 2D slices from just one 3D PET dataset, while our $|S|=4$ or 8 cases sampled slices from 495 distinct 3D PET datasets. Despite this, the $|S|=12$ case still offered visually superior results to OSEM. This case in particular may benefit from randomly weighting summed images for greater diversity.

In Figure \ref{fig:mean_roi}, we further analyse our method, and see that for $|S|<12$ it offers lower background noise for a given gray-to-white matter contrast than OSEM, MAP-EM and reconstruction using an SGM trained on real PET data.

While our method uses MR information, it should be noted that this is only as training data for registration, and not for data conditioning in the iterative reconstruction process. 

\vspace{-0.2cm}
\section{Summary}

We have proposed principled new methodologies for generating diverse and high-quality pseudo-PET images and applying them for SGM-based PET reconstruction. Our results show that our pseudo-PET images are useful for learning highly subject-specific priors for PET image reconstruction. Further research is needed to compare our methodology to standard MR-guided reconstructions and using MR-guided reconstruction as training data for an SGM.

\vspace{-0.2cm}
\appendices

\bibliography{IEEE_MIC_Abstract_2_refs}

\end{document}